\documentclass[12pt]{aastex}
\usepackage{emulateapj5}
\usepackage{endfloat}
\nofiglist
\notablist
\nomarkersintext
\def\txtfig#1{}

\def\begfig{\begin{figure*}}
\def\endfig{\end{figure*}}
\usepackage{amsmath}
\usepackage{amssymb}
\usepackage{graphicx}
\def\revone#1{{#1}}
\def\revrm#1{}
\def\revtwo#1{{#1}}
\def\CH#1{#1}

\def\kel{\hbox{K}}

\def\AU{\hbox{AU}}

\def\comma{\,,}
\def\fullstop{\,.}
\def\eff{\hbox{{\scriptsize eff}}}
\def\erf{\mathrm{erf}}
\def\innn{{\mathrm{rim}}}
\def\outtt{\hbox{{\scriptsize out}}}
\def\evap{\mathrm{evap}}

\def\irr{\mathrm{irr}}
\def\emit{\mathrm{emit}}
\def\abs{\mathrm{abs}}
\def\dlt{\delta}
\def\deg{o}
\def\iincl{i}
\def\Hsin{H_{\mathrm{rim}}}
\def\Hpin{h_{\mathrm{rim}}}

\def\HsCG{H_{\mathrm{cg}}}
\def\HpCG{h_{\mathrm{cg}}}

\def\chiin{{\chi_{\mathrm{rim}}}}
\def\chiCG{{\chi_{\mathrm{cg}}}}
\def\Hr{H_{\mathrm{rim}}}
\def\Ti{T_{\mathrm{i}}}
\def\Ts{T_{\mathrm{s}}}
\def\trim{T_{\mathrm{rim}}}
\def\tdisk{T_{\mathrm{cg}}}
\def\Hs{H}
\def\Hp{h}
\def\rinn{R_{\mathrm{rim}}}
\def\surf{\mathrm{s}}
\def\interior{\mathrm{i}}

\def\rim{\mathrm{\innn}}

\def\radial{\mathrm{r}}
\def\vert{\mathrm{v}}
\def\rflare{R_{\mathrm{fl}}}
\def\rout{R_{\mathrm{out}}}
\def\fli{\mathrm{i}}
\def\fls{\mathrm{s}}

\newcommand{\rnir} {$\frac{F_{NIR}}{F_{*}}$}

\newcommand {\fsil}{$\frac{F_{10}}{F_{7.7}}$}

\newcommand{\simless}{\mathbin{\lower 3pt\hbox
      {$\rlap{\raise 5pt\hbox{$\char'074$}}\mathchar"7218$}}} 
\newcommand{\simgreat}{\mathbin{\lower 3pt\hbox
     {$\rlap{\raise 5pt\hbox{$\char'076$}}\mathchar"7218$}}} 
\newcommand{\um} {$\mu$m}
\newcommand{\nuFnu} {$\nu$F$_\nu$}
\begin{document}
\title{Passive irradiated circumstellar disks with an inner hole} 
%
%
\author{C.P.~Dullemond}
\affil{Max Planck Institut f\"ur Astrophysik,\\
Postfach 1317, D--85741 Garching, Germany\\
e--mail: dullemon@mpa-garching.mpg.de}
\and
\author{C.~Dominik}
\affil{Sterrenkundig Instituut `Anton Pannekoek',\\
Kruislaan 403, 1098 SJ Amsterdam, The Netherlands \\
e--mail: dominik@astro.uva.nl}
\and
\author{A.~Natta}
\affil{Osservatorio Astrofisico di Arcetri, Largo E.Fermi 5,\\
 I-50125 Firenze, Italy\\
e--mail: natta@arcetri.astro.it}

%
%
%
%

%
%
\begin{abstract}
%
%
%
%
  A model for irradiated dust disks around Herbig Ae stars is proposed. The
  model is based on the flaring disk model of Chiang \& Goldreich (1997,
  henceforth CG97), but with the central regions of the disk removed.  The
  inner rim of the disk is puffed up and is much hotter than the rest of the
  disk, because it is directly exposed to the stellar flux. If located at
  the dust evaporation radius, its reemitted flux produces a conspicuous
  bump in the SED which peaks at 2--3 micron. We propose that this emission
  is the explanation for the near-infrared bump observed in the SEDs of
  Herbig Ae stars. We study for which stellar parameters this bump would be
  observable, and find that it is the case for Herbig Ae stellar parameters
  but not for T-Tauri stars, confirming what is found from the
  observations. We also study the effects of the shadow cast by the inner
  rim over the rest of the flaring disk. The shadowed region can be quite
  large, and under some circumstances the entire disk may lie in the shadow.
  This shadowed region will be much cooler than \revone{an unshadowed
  flaring} disk, since its only heating sources are radial radiative
  diffusion and possible indirect sources of irradiation. Under certain
  special circumstances the shadowing effect can suppress, or even
  completely eliminate, the 10 micron emission feature from the spectrum,
  which might explain the anomalous SEDs of some isolated Herbig Ae stars in
  the sample of Meeus et al.~(2001).  At much larger radii the disk emerges
  from the shadow, and continues \revone{as a flaring disk} towards the
  outer edge. The emission from the inner rim contributes significantly to
  the irradiation of this flaring disk. The complete semi-analytical model,
  including structure of the inner edge, shadowed region and the flared
  outer part, is described in detail in this paper, and we show examples of
  the general behavior of the model for varying parameters.
%
%
\end{abstract}
%
%
%
%

\section{Introduction}
The hypothesis that Herbig Ae/Be stars (HAeBe; Herbig 1960) are intermediate
mass pre-main-sequence stars is today generally accepted. \revone{The large
infrared (IR) excess observed from these stars can be naturally explained in
this context by emission from a protostellar/protoplanetary disk. This
picture} has been given more credibility in recent years by interferometric
observations at millimeter wavelengths (Mannings \& Sargent 1997,
2000). \revone{Over the years, various disk models have been proposed for
pre-main-sequence stars, among which are disks powered by accretion (Lin \&
Papaloizou 1980, Bell \& Lin 1994), irradiated non-accreting (passive) disks
(Kenyon \& Hartmann 1987, Chiang \& Goldreich 1997 henceforth CG97), and
irradiated accretion disks (Ruden \& Pollack 1991, Calvet et al.~1991, 1992,
D'Alessio et al.~1998, 1999, Bell 1999). The spectral energy distributions
(SEDs) of Herbig Ae stars seem to be best fitted by passive irradiated
flaring disks, at least at mid- and far-infrared wavelengths (Chiang et
al.~2001). At near-IR wavelengths, however, the flaring disk picture fails
to agree with the observed SEDs. The same problem is encountered when
fitting non-irradiated accretion disk models to the SEDs. Hillenbrand et
al.~(1992) noticed that in 29 out of 51 HAeBe stars (their Group I sources)
the SED at near and mid-IR wavelengths was well fit by models of
geometrically flat, optically thick disks with very large accretion rates,
but that these disks needed to be optically thin in their inner parts, to
account for the distinctive inflection shown by all HAeBe stars at
$1.2\simless \lambda \simless 2.2$ \um\ (see also Lada \& Adams
1992). Better measured SEDs for a number of Herbig AeBe stars of spectral
type A (HAe in the following) have confirmed the Hillenbrand et al.~(1992)
results.}

All the observed \revone{isolated} HAe stars
have large near-infrared emission in excess of the photospheric one with
very similar dependence on wavelength, that can be described as a
near-infrared ``bump'': the excess monochromatic luminosity \nuFnu\ is
negligible around 1 \um, peaks around 2 \um\ and decreases slowly with
wavelength to about 8\um, where the silicate emission feature starts (Meeus
et al. 2001; Natta et al. 2001). These stars have very little
extinction, so that reddening in the near-infrared is negligible.

The negligible opacity of the inner disk region, required to explain the
near-infrared bump is not consistent with the high accretion rates necessary
to account for the large near-infrared excess, as discussed by Hartmann,
Kenyon \& Calvet (1993). These authors suggested that, given the
difficulties of applying disk models to HAeBe stars, one should consider the
hypothesis that the emission originates in a dusty \revone{circumstellar
envelope, possibly containing} very small grains transiently heated by
ultraviolet photons. This possibility, however, has been ruled out by
theoretical models (Natta \& Kr\"ugel 1995) and by ISO observations (Meeus
et al. 2001).

An alternative explanation for the near-infrared behaviour of HAe stars has
recently been proposed by Natta et al.~(2001). \revone{Their model starts
from the notion that the mid- and far-infrared excess of HAe stars can be
well fitted with a passive irradiated flaring disk, as was mentioned above.}
%
%
As expected, these models fail to explain the observations in the
near-infrared. However, Natta et al.~have noticed that, if one truncates the
\revone{flaring} disk at a certain radius (i.e.~introduce a hole),
\revone{the emission from the inner rim of this disk is
non-negligible}. This inner rim will have a \revone{considerable} covering
fraction around the star, and it will acquire a temperature much hotter than
what the \revone{flaring} disk interior at the same radius would be. This is
because the \revone{flaring} disk receives the stellar radiation under a
small grazing angle, and will therefore be relatively cool, while the inner
rim is in full sight of the star and will therefore be hot. In fact, if one
assumes that the inner edge of the disk is naturally located at the radius
where dust grains acquire their evaporation temperature, then the emission
from this inner rim is going to produce a near-infrared bump very
reminiscent of the observed one in shape and intensity.

In this paper we wish to proceed further along this path, and investigate a
number of issues that have not been addressed by Natta et al. (2001).
Firstly, we will formalize the description of the structure of a disk with a
large inner radius, using the \revone{flaring disk models of} CG97 as a
starting point.  In particular, we wish to investigate how the presence of
the hot inner rim modifies structure and appearance of the rest of the
disk. Since the inner rim is much hotter than the CG97 disk at the same
radius, its vertical scale height will be larger as well. In other words,
the inner rim of the disk will puff-up. This will enhance the near-infrared
emission, but also cast a shadow over the disk behind it. Part of the disk
which would normally be directly illuminated by the stellar radiation, will
now be in the shadow, and therefore collapse and cease to emit IR radiation.
At large enough radii, it is possible that the flaring shape of the CG97
disks will be conserved, and that the disk will grow \revone{out of} the
shadow, unchanged by the presence of the rim.  We thus propose a scenario in
which the emission from the reprocessing disk comes from a hot inner rim,
caused by dust evaporation, plus the non-shadowed part of a truncated CG97
disk.

A second point we will discuss concerns the behaviour of the near-infrared
emission as function of the spectral type of the star.  The IR spectra of T
Tauri stars (TTS) are well described by disk models with small or zero inner
holes (Beckwith et al. 1990; Kenyon, Yi \& Hartmann 1996) and moderate
accretion rates (Gullbring et al. 1998). The few known pre-main-sequence
stars of spectral type F and G seem closer in properties to TTS than to HAe
stars (Hillenbrand et al. 1992; Hartmann et al. 1993).  We show that this
behaviour can be explained by the same disk models and that it is caused by
a combination of effects. First of all, the relative importance of the NIR
bump emission from the inner edge in comparison to the disk emission is
smaller for stars of later spectral type. Secondly, the stellar atmospheric
spectrum shifts towards longer wavelengths as the effective temperature of
the star decreases, swamping the rim emission. Finally, it is possible that
TTS have on average higher accretion rates, and therefore more optically
thick inner disks, than HAe stars.

\revtwo{The structure of the paper is as follows. Section 2 discusses the
equations, Section 3 describes how the SED is computed. The results will be
presented in Section 4, and we conclude with a discussion in Section 5.}

\section{The structure of a disk with a large inner hole}

In this section we will derive the equations governing the structure
of a passively irradiated circumstellar disk with an inner hole.  The
main effect of the inner hole on the disk structure is a vertical
boundary which is irradiated directly by the star, heating it to
temperatures higher than what would be expected at the same location
for a thin disk with grazing irradiation only.  Because of the higher
temperature, the disk boundary will be puffed up, and the material
immediately outside the inner rim will be shadowed, leading to a
decreasing scale height.  At large distances from the star it
is possible that the disk becomes high enough to emerge from the shadow
of the inner rim, in which case it may continue to flare like a disk
without a puffed-up inner rim.  However, it is not certain that the
disk will be able to leave the shadow.  This is a dynamical question
and the answer may depend upon the history of the disk and many other
parameters (e.g.~was the disk flaring when the inner boundary started
to emerge and to puff up?).

\begin{figure}
\plotone{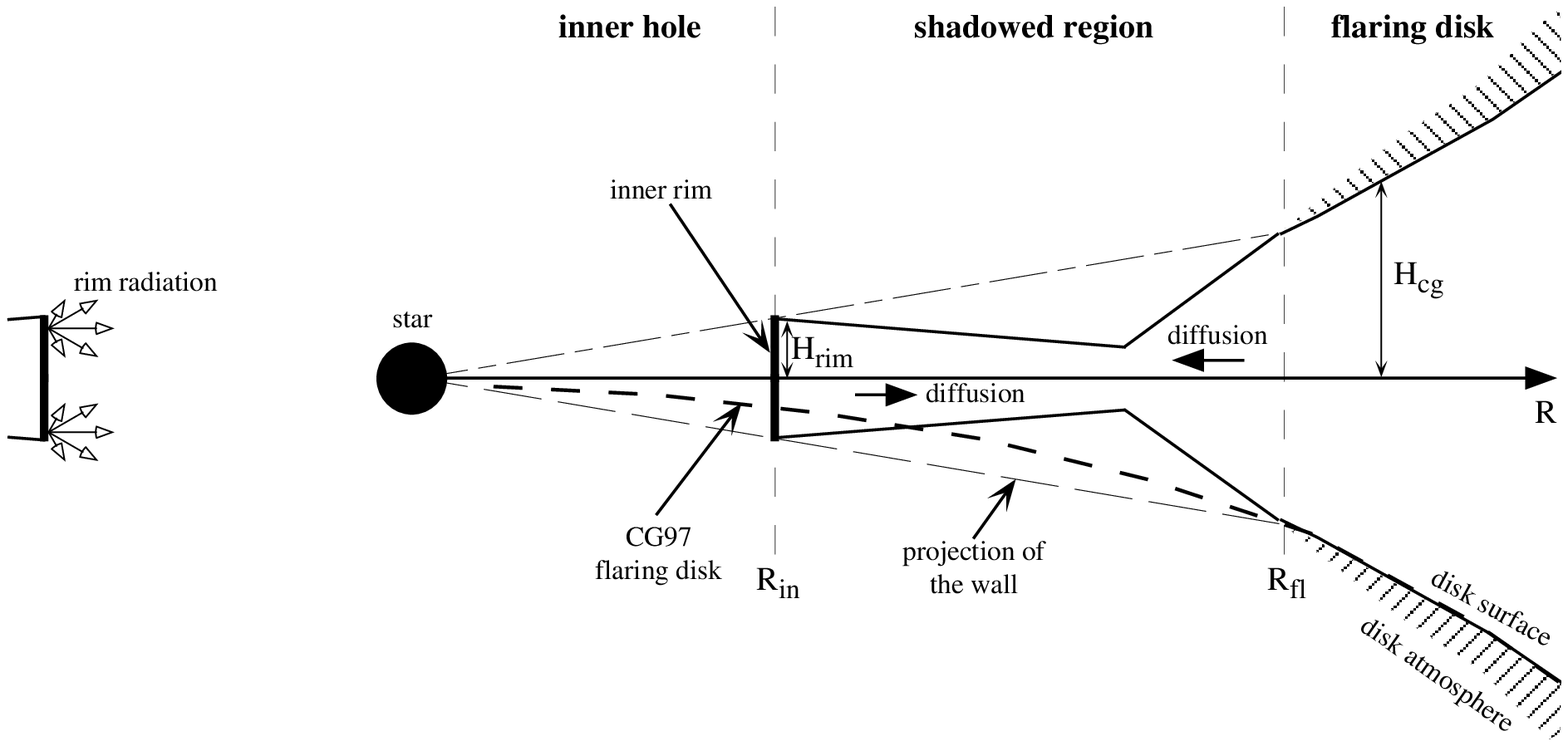}
\caption{A sketch of the geometry of the model.}
\label{fig-disk-hole-geometry}
\end{figure}

Figure \ref{fig-disk-hole-geometry} shows our scenario, where we assume that
the flared part of the disk exists when it can. 

We will divide the disk structure calculations in three parts (inner rim,
shadowed region, and flared component) and derive the basic equations
governing the structure in the three regions.  We start with the well-known
flared part.

\subsection{Flaring disk}\label{sec:flare-disk}

For the the flaring part of the disk we follow the line of Chiang \&
Goldreich (CG97) with some minor modifications.  For completeness, and in
order to introduce the nomenclature we repeat the basic equations without
physical discussion. \revrm{For details see CG97.}

\subsubsection{Disk interior}
Direct stellar radiation impinges onto the disk at an angle $\alpha$ given
by $\alpha = 0.4 R_{*}/R + R\,d(\HsCG/R)/dR$, where $R_{*}$ is the radius of
the star, $R$ the distance from the star in the disk plane and $\HsCG$ the
height of the disk photosphere above the midplane.  Anticipating a nearly
powerlaw behavior, we replace $Rd/dR$ \revone{by $\gamma-1$, where $\gamma$
is a slowly varying function of the order of $9/7$}. We can thus rewrite the
expression for $\alpha$ as:
\begin{equation}\label{eq-cgalpha}
\alpha = \frac{0.4 R_{*}}{R} + (\gamma-1) \frac{\HsCG}{R}
\fullstop
\end{equation}
\revtwo{The $(\gamma-1)$ factor is self-consistently determined as a
function of $R$, using a numerical method devised by Chiang et al.~(2001).
In this way energy conservation is ensured, while the assumption of a
constant $(\gamma-1)$ can cause deviations by a factor $\sim 2$.}
The stellar flux impinging with this angle into the disk is:
\begin{equation}\label{eq-irrad-flux-star}
F^{*}_{\irr}=\alpha\frac{L_{*}}{4\pi R^2}
\fullstop
\end{equation}
This equation assumes that the entire star is visible from the
considered point on the disk surface.  If the disk would stretch all
the way to the star, the bottom half of the star would not be visible
and a correction would have to be made.  However, since we consider
disks with large inner holes, the transition from the star being invisible
to the star being fully visible will be sharp, and we ignore this effect.

\revone{The flux $F^{*}_{\irr}$ is absorbed in the uppers layer of the disk,
which will re-radiate half of the flux away from the disk, and half down
into the disk's deeper layers. A fraction $\psi_{\surf}$ of this flux will
be absorbed by the interior, where $\psi_{\surf}$ is a dimensionless number
in between $0$ and $1$ that accounts for the possibility that the disk
interior is not fully optically thick to the emission of the surface
layer. We assume the disk interior to be isothermal with a temperature
$\Ti$, which is a very good approximation for optically thick disks. This
interior will emit a flux $F_{\emit}$ given by
\begin{equation}\label{eq-emit-flux-interior}
F_{\emit}=\psi_{\interior}\sigma\Ti^4
\fullstop
\end{equation}
where $\psi_{\interior}$ accounts for the possibility that the disk
interior is not fully optically thick to its own emission. For a
fully optically thick disk one has $\psi_{\surf}=\psi_{\interior}=1$.
The formulas for determining $\psi_{\surf}$ and $\psi_{\interior}$ for
not optically thick cases will be given in appendix
\ref{app:subsec-cg-finetune-psi}.}

\revone{By equating $(1/2)\psi_{\surf}F^{*}_{\irr}=F_{\emit}$, one can solve
for $\Ti$ and find:}
\begin{equation}\label{eq-cgeq-tempi}
\Ti = \left(\frac{\alpha\psi_{\surf}}{2\psi_{\interior}}\right)^{1/4}\;
\left(\frac{R_{*}}{R}\right)^{1/2}T_{*} 
\fullstop
\end{equation}

The disk will be almost isothermal in the vertical direction, leading
to a Gaussian density distribution with a constant pressure scale
height $\HpCG$ given by
\begin{equation}\label{eq-vert-press-bal-cg}
\frac{\HpCG}{R} = \left(\frac{\Ti}{T_c}\right)^{1/2}
\left(\frac{R}{R_{*}}\right)^{1/2}
\end{equation}
with $T_c=GM_{*}\mu m_p/k R_{*}$ is the virial temperature at the stellar
surface. The ratio of the disk surface height $\HsCG$ to the pressure scale 
height $\HpCG$ is a dimensionless number $\chiCG$ of order unity:
\begin{equation}
\HsCG = \chiCG \HpCG
\end{equation}
Usually this $\chiCG$ will lie somewhere in between $2$ and $6$. But it can be
self-consistently determined from the grazing angle $\alpha$, the surface
density $\Sigma$ and the Planck mean opacity at the stellar temperature
$\kappa_P(T_{*})$. The way to do this will be described in appendix
\ref{app:subsec-cg-finetune-chi}.

\subsubsection{\CH{Disk \revone{surface layer}}}
\label{subsec-surfacelayer}
\CH{Since the disk \revone{surface layer}} is defined as the layer of
matter in direct sight of the central star up to grazing optical depth
$\tau_{\alpha}\simeq 1$, the temperature in this layer can be
estimated from the optically thin expression:
\begin{equation}\label{eq-surface-layer-temp}
\Ts= \frac{1}{\epsilon_s^{1/4}}\left(\frac{R_{*}}{2R}\right)^{1/2}T_{*}
\comma
\end{equation}
where $\epsilon_s$ is the ratio of the Planck mean opacities at $\Ts$ and
$T_{*}$. \CH{Quantities describing the surface layer carry the subscript
$s$}. The Planck mean opacity is defined as
\begin{equation}
\kappa_P(T)=\frac{\int_0^\infty B_\nu(T)\kappa_\nu d\nu}
{\int_0^\infty B_\nu(T) d\nu}
\fullstop
\end{equation}
The flux from the surface layer in both the upwards and the downwards
direction is:
\begin{equation}
\begin{split}
F_{s} &= 2\pi\Delta\Sigma\int_0^\infty B_{\nu}(\Ts)\kappa_\nu d\nu \\
&= 2\Delta\Sigma\kappa_P(\Ts)\sigma \Ts^4 
\comma
\end{split}
\end{equation}
where $\Delta\Sigma$ is the surface density of the layer.  Energy
conservation implies $F_s=(\psi_{\interior}/\psi_{\surf})\sigma
\Ti^4$, so the surface density of the layer is:
\begin{equation}\label{eq-delta-sigma}
\Delta\Sigma = \frac{1}{2\kappa_P(\Ts)}\,
\frac{\psi_{\interior}}{\psi_{\surf}}\, \frac{\Ti^4}{\Ts^4}
\fullstop
\end{equation}

\revone{In our treatment of the surface layer we ignore scattering. It
should be pointed out that, while the effect of this approximation on the
structure of the disk is small, the effect on the SED might be
non-negligible.}

\subsection{Inner rim}
We assume in the following that the disk is truncated on the inside by dust
evaporation. The disk may continue towards the star in purely gaseous form,
but, as long as the gas is optically thin to the stellar radiation, we can
neglect its effects and describe the disk structure as if it has a true
inner hole.

The inner disk radius $\rinn$ is determined by requiring that the disk
temperature equals the dust evaporation temperature, which we take
$T_{\evap}=1500\kel$. The inner rim of the disk exposes a vertical surface
to direct stellar radiation and will therefore be hotter than \revone{flaring} disk
model at the same radius. For a given inner radius $\rinn$ of the disk, the
blackbody temperature of the rim is:
\begin{equation}\label{eq-trim-given-rin}
\trim = \left(\frac{L_{*}}{4\pi \rinn^2 \sigma}\right)^{1/4}\;
(1+\Hr/\rinn)^{1/4}
\comma
\end{equation}
which is derived by equating the emitted blackbody flux $\sigma T_{\innn}^4$
to the received flux from direct stellar radiation $L_{*}/4\pi \rinn^2$.
The factor $(1+\Hr/\rinn)^{1/4}$ approximately accounts for the effect of
self-irradiation, i.e.~irradiation of the rim by its own emission.
\revone{The ratio $\Hr/\rinn$ is an approximate expression for the covering
fraction of the rim enclosing the cavity $R<\rinn$.} \CH{We assume $R_{*}\ll
R_{\innn}$ and therefore neglect that the star itself obscures a small
region on the far side of the rim.}  The vertical height of the inner rim,
$\Hr$, is given by
\begin{equation}\label{eq-hsrim-from-hprim}
\Hr=\chiin\Hpin
\comma
\end{equation}
where $\Hpin$ is the pressure scale height at the inner rim
\begin{equation}\label{eq-hprim-from-press}
\Hpin=\sqrt{\frac{k\trim\rinn^3}{\mu m_p G M_{*}}}
\comma
\end{equation}
\CH{and $\chiin$ is a dimensionless constant (similar to $\chiCG$ defined
above).  In appendix \ref{app:sec:surface-height-at-inner-rim} we show how
to compute $\chiin$ self-consistently.}

By equating $\trim=T_{\evap}\equiv 1500\kel$ one can solve for $\rinn$ by
writing Eq.(\ref{eq-trim-given-rin}) as
\begin{equation}
\rinn = \left(\frac{L_{*}}{4\pi \trim^4 \sigma}\right)^{1/2}\;
(1+\Hr/\rinn)^{1/2}
\comma
\end{equation}
and iterate until convergence. If the self-irradiation can be ignored
($\Hr/\rinn\ll 1$), then no iteration is needed.  However, as it turns out,
for most HAeBe stars the ratio $\Hr/\rinn$ (which is the covering fraction
of the inner rim) is of the order of $0.1$ to $0.25$, and therefore the
effects of self-irradiation cannot be ignored.

\CH{Self-irradiation has the effect of pushing out the inner radius, if the
temperature of the inner rim is fixed at $1500\kel$. This means that the
blackbody-emitting surface is larger. But since only part of this radiation
will escape to infinity (the other part being again absorbed by the inner
rim itself), energy remains conserved.  Once $\rinn$ is determined, the
vertical height of the inner rim $\Hr$ automatically follows from
Eqs.(\ref{eq-hsrim-from-hprim},\ref{eq-hprim-from-press}), and one has the
basic properties of the inner rim fixed.}

\subsection{Shadowed region}
\label{sec:shadowed-region}
The inner rim will always be significantly higher than the
\revone{flaring} disk height, because it is hotter by a factor of at
least $(2/\alpha)^{1/4}$. In general this will roughly yield a
puffing-up ratio of $\Hsin/\HsCG\simeq 2$. The disk material
immediately behind this inner rim will therefore be deprived of direct
stellar light, since the rim casts a shadow over the disk behind it.
This region of the disk is cooler than predicted by CG97 models, and
\CH{therefore not flared}.  Only at much larger distance, where the
disk comes out of the inner rim shadow and is illuminated again by the
stellar light, flaring \revrm{as predicted by CG97} may occur.  Since the
shadow covers every point $(R,z)$ for which $z/R<\Hsin/\rinn$, the
radius $\rflare$ at which the disk can again be described by the CG97
model is computed by solving the following equation:
\begin{equation}\label{eq:big-shadow}
\Hsin(\rinn) \frac{\rflare}{\rinn} = \HsCG(\rflare)
\fullstop
\end{equation}

For $R<\rflare$, the most important sources of irradiation are the
near-infrared emission of the rim itself and the fraction of stellar light
that is scattered toward the disk by the edge of the rim, or by a halo of
dust surrounding the star+disk system (Natta 1994).  In addition to this,
radiative diffusion can exchange energy between neigboring annuli.  However,
rough estimates of these effects show that the radiation emitted by this
shadowed region of the disk is negligible compared to the emission of the
rim and of the outer disk, directly heated by the star.

\begin{figure}
\epsscale{0.5}
\plotone{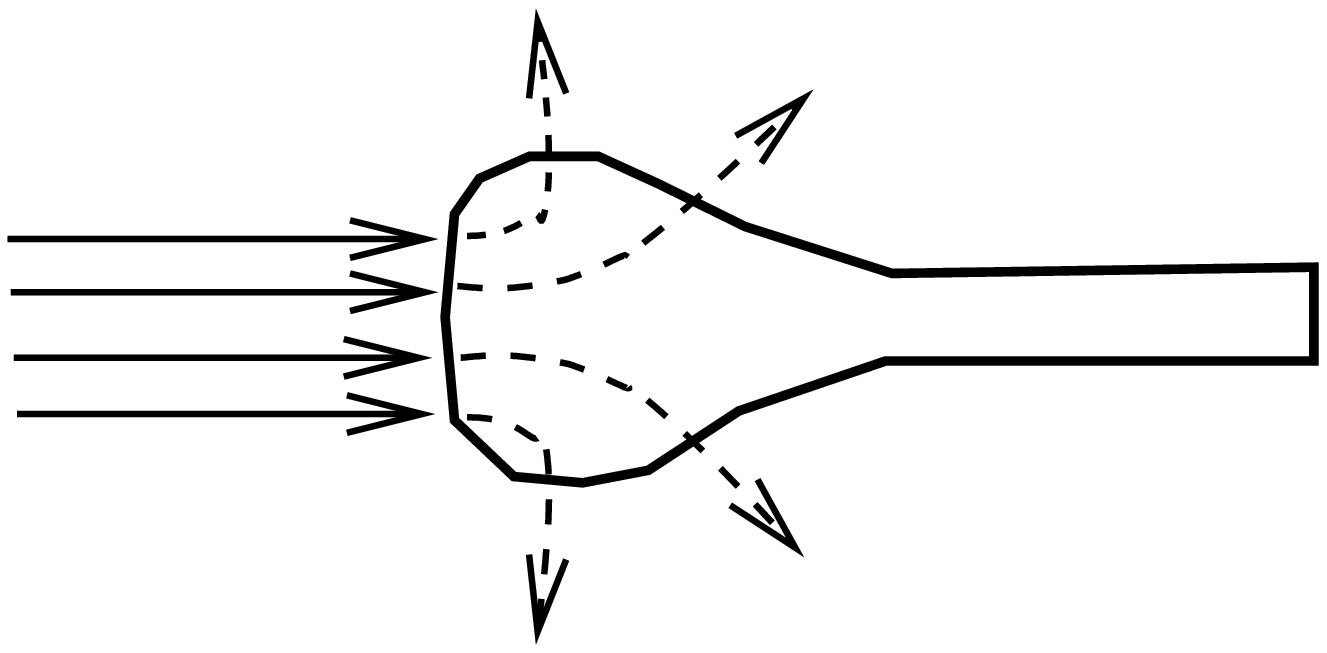}
\epsscale{1.0}
\caption{A sketch of the two dimensional radiative diffusion 
process through the disk inner edge. \revtwo{The solid arrows indicate
direct radiation from the star. The dashed arrows represent the 
diffusion through the disk.}}
\label{fig-diffusion-rim}
\end{figure}

However, the disk structure immediately behind the inner rim is of
importance for our discussion, because we need it in order to estimate the
quantity $\chiin$ in Eq.(12), i.e.~the ratio of the photospheric to the
pressure scale height at $\rinn$ (see
Appendix~\ref{app:sec:surface-height-at-inner-rim}). At this location of the
disk none of the irradiation sources play any significant role.  Instead,
direct radiative diffusion from the inner rim will dominate the energy
balance there (see Fig.~\ref{fig-diffusion-rim}).
It is difficult to determine exactly what the structure of the disk is in
this diffusive region, but a simple model for this will be provided below.

The inner rim of the diffusive region receives direct starlight. This
radiation is absorbed and re-emitted by the dust grains. Some of this
radiation will once again be absorbed and re-emitted. In the end, most
radiation will, after one or more absorption/re-emission events escape from
the hot inner rim surface. But a small fraction of the radiation remains
trapped within the disk and diffuses towards larger radii. The diffusive
flux is proportional to the gradient of the temperature of the disk's
interior. Since it is expected that the temperature has a net decrease
outwards, there will be an outward pointing diffusive flux. As a package of
energy has moved one \CH{surface height $\Hs$} towards larger radii, it has
had an equal chance of having escaped from the disk's surface. This is true
for each surface height it travels outwards. The differential equation
for the disk's temperature would be \revone{approximately}:
\begin{equation}
\label{eq:1}
\frac{d(RT^4)}{dR} \simeq -\frac{RT^4}{\Hs}
\comma
\end{equation}
\revone{where $T$ is a rough estimate of the temperature of the disk
interior at radius $R$. In this equation we have included the effect of
geometric dilution due to the cylindrical geometry, hence the factor $R$
inside the brackets of the differentiation. A similar factor $\Hs$ is not
included, since $\Hs$ is expected to decrease outwards, which would lead to
an unphysical geometrical compression of the radiation field. In this
diffusive region the disk could be vertically non-isotermal, but we ignore
this effect in this simple treatment. Also we ignore the fact that the
definition of $\Hs$, as given in Section \ref{sec:flare-disk} is not very
meaningful in this non-irradiated region.} \revone{Now, since $\Hs \propto
\sqrt{TR^3}$ (see e.g.~Eq.\ref{eq-vert-press-bal-cg}, and assume for
convenience that $\chi=$const) the temperature $T$ can be eliminated from
the differential equation, leaving an ordinary differential equation to be
solved for $\Hs$:}
\begin{equation}\label{eq-full-diff}
\frac{d\lg(\Hs^8R^{-11})}{dR}=-\frac{1}{\Hs}
\fullstop
\end{equation}
This equation can be solved numerically.
\revtwo{For $\Hs\ll R$, it reduces to
\begin{equation}
\frac{d(\Hs/R)}{dR}=-\frac{1}{8R}
\comma
\end{equation}
which shows that at the inner edge $\Hs/R$ approximately goes as
$\Hs/R=\Hs(\rinn)/\rinn-\lg(R/\rinn)/8$.} 
%
%
\revtwo{In this approximation the disk has lost all its energy at
$R-\rinn=8H(\rinn)$, and has collapsed to zero height.} This would be of
course unphysical, and before this happens, other heating mechanisms will
have taken over, such as the ring emission mentioned above, light scattering
off a spherical halo surrounding the disk+star system, interstellar
radiation or viscous dissipation.

\revtwo{Eq.(\ref{eq-full-diff}), but with an opposite sign,} describes the
diffusion of energy inwards from the still flaring part of the disk into the
shadowed region.

\subsection{Self-irradiation corrections to flaring part}
\label{sec:self-irrad-corr}

\CH{For $R>\rflare$ the flaring disk is in full view of the central
  star, and the irradiation is clearly dominated by the stellar flux.
  However, for HAeBe stars the luminosity of the inner rim can be of
  order $20\%$ of the stellar flux.  Moreover, the emission of the
  inner rim is not isotropic.  In close to polar directions, the
  projected surface of the rim is small in the flux correspondingly
  low.  For a point on the surface of the flaring disk on the other
  hand, the projected surface of the rim is close to
  $4R_{\innn}H_{\innn}$ (see Appendix~\ref{app:projsurf}).  Therefore,
  the relative flux contribution of the rim at the disk surface will even be
  larger than the angular average of 20\%.}  For an accurate
determination of the disk structure in the flared region one should
include this secondary irradiation. We consider the
inner rim as a cylinder. The irradiative flux is then (see appendix
\ref{app:projsurf}):
\begin{equation}
\begin{split}
F_{\irr} = & \alpha(R)F_{*} + \frac{2}{\pi}\alpha(R)\sigma \trim^4 
\left(\frac{\rinn}{R}\right)^2 \\ 
&\cos \theta\; [\delta\sqrt{1-\delta}+\arcsin\delta]
\fullstop
\end{split}
\end{equation}
\revone{This irradiative flux $F^{\irr}_{*}$ is then} equated to the 
disk's reemission flux $F_{\emit}$ (Eq.\ref{eq-emit-flux-interior}),
thus yielding the corrected value of $\Ti$.

\revone{For the surface layer temperature one also has to add the two sources
of irradiation. But in order to compute the $\epsilon_s$ factor in
Eq.(\ref{eq-surface-layer-temp}), one now has to replace the
$\kappa_P(T_{*})$ in the definition of $\epsilon_s$ (Section
\ref{subsec-surfacelayer}) with:
\begin{equation}
\label{eq:stelplusrim}
\kappa_{\irr} \equiv \frac{\int_{0}^{\infty} \kappa_\nu F^{\irr}_\nu
  d\nu}{\int_{0}^{\infty} F^{\irr}_\nu d\nu}
\end{equation}
}

\section{Computing the SED}
Once the disk structure is known, one can compute the emerging spectrum.  In
principle one requires a 2-D (axially symmetric) ray-tracing program to
compute the SED at various inclination angles. This is because the disk
structure, in particular for HAeBe stars, is far from being flat, and
self-occultation is a real possiblity (Chiang \& Goldreich 1999). Yet, in
this section we shall use a simplified treatment of
inclination, therefore limiting ourselves to inclinations that are not too
edge-on.

\subsection{Spectrum of flaring disk}
The determination of the spectral components from the flaring part of
the disk is relatively straightforward, and has been described fully
by CG97. We \revrm{refer to their paper for the details, and} confine
ourselves here to a quick reference of the equations. The emission
from the interior of the disk is:
\begin{equation}
\begin{split}
F_\nu^{\fli}(d,\iincl) = 2\pi \cos\iincl  \int_{\rflare}^{\rout} &
[1-\exp(-\Sigma(R)\kappa_\nu/\cos \iincl)]\\
&\times B_\nu(\Ti(R))\frac{R}{d^2}\,dR
\comma
\end{split}
\end{equation}
where $d$ is the distance from the observer to the source, and $\iincl$ the
inclination angle ($\iincl=0$ means face-on). This equation remains valid at
wavelengths where the disk interior becomes optically thin. The emission
from the surface layer is
\begin{equation}\label{eq-fnu-surf}
\begin{split}
F_\nu^{\fls}(d,\iincl) = 2\pi \int_{\rflare}^{\rout} &
[1+\exp(-\Sigma(R)\kappa_\nu/\cos \iincl)]\\
&\times B_\nu(\Ts(R)) \Delta\Sigma \kappa_\nu\frac{R}{d^2}\,dR
\comma
\end{split}
\end{equation}
\revone{where $\Delta\Sigma$ is given by Eq.(\ref{eq-delta-sigma}).
Eq.(\ref{eq-fnu-surf}) includes the effect of also seeing the surface layer
on the other side of the disk at wavelengths where the disk is optically
thin.}

\subsection{Spectrum of inner rim}
The inner rim is assumed to be a cylinder emitting on its inner side as a
blackbody with temperature $\trim$. To compute the emission from this rim we
have to include inclination effects. Clearly, if $\iincl=0$ (face-on), then
the emission from the rim is zero (although in reality the rim may be more
rounded-off, and therefore also visible at face-on inclination). For
$\iincl=\pi/2$ (edge-on) the emission is also zero, because the rim is
self-occulting. The projected surface, as seen by an observer at inclination
$\iincl$ is given by Eq.(\ref{eq-flux-from-rim-incl-dsmall}).  For
not-too-large inclination angles we are in the regime $\delta\ll 1$ and one
can then write:
\begin{equation}
F_\nu^{\rim}(d,\iincl) = 4 \frac{\rinn\Hsin}{d^2} \sin \iincl B_{\nu}(\trim)
\fullstop
\end{equation}

The total luminosity of the rim equals $L_{\rim} = \Omega_{\rim} L_{*}$,
where $\Omega_{\rim}=\Hsin/\rinn$ is the covering factor of the rim
(i.e.~the fraction of the sky covered by the rim as seen from the
perspective of the star). For large $\Hsin/\rinn$ this will therefore give a
strong contribution to the SED.

\def\NIR{\mathrm{NIR}}
\def\IR{\mathrm{IR}}

\begin{table*}
\caption{Properties of different models discussed in this paper.}
\revone{%
\centerline{%
\begin{tabular}{lcc|rrccrrrr|rrrrr}
Series & Fig & Model & $T_{\eff}$ & $\frac{F_{*}}{F_{\odot}}$ & $\frac{\Sigma_{0}}{\mathrm{g\,cm}^{-2}}$ & $\beta$ & $\frac{\rinn}{\mathrm{AU}}$ & $\frac{\rflare}{\mathrm{AU}}$ &$\frac{\Hsin}{\rinn}$ & $\chi_{\innn}$ & $\frac{F_{\mathrm{NIR}}}{F_{*}}$
&  $\frac{F_{\mathrm{IR}}}{F_{*}}$ & $\frac{F_{\mathrm{NIR}}}{F_{\mathrm{IR}}}$ & $\frac{F_{10}}{F_{7.7}}$ \\[1mm]\hline
Example&
\ref{fig-stmodel-sed-all} &
      E1 &  9500 &   47    & 2000 & -1.5 & 0.47  &  0.0   & 0.11  & 5.3  & 0.03  & 0.35  & 0.08  & 4.7 \\
   && E2 &  9500 &   47    & 2000 & -1.5 & 0.50  &  0.0   & 0.10  & 2.9  & 0.12  & 0.50  & 0.24  & 3.6 \\
   && E3 &  9500 &   47    & 2000 & -1.5 & 0.52  &  6.6   & 0.18  & 5.0  & 0.17  & 0.50  & 0.33  & 3.5 \\\hline
Inclination &
\ref{fig-stmodel-effect-all}A &
      I1(50$^{o}$) &  9500 &   47    & 2000 & -1.5 & 0.52  &  6.0   & 0.18  & 5.0  & 0.18  & 0.50  & 0.36  & 3.2 \\
   && I2(35$^{o}$) &  9500 &   47    & 2000 & -1.5 & 0.52  &  6.0   & 0.18  & 5.0  & 0.13  & 0.47  & 0.28  & 3.9 \\
   && I3(20$^{o}$) &  9500 &   47    & 2000 & -1.5 & 0.52  &  6.0   & 0.18  & 5.0  & 0.08  & 0.43  & 0.19  & 5.2 \\\hline
$R_{\innn}$ &
\ref{fig-stmodel-effect-all}B &
      R1 &  9500 &   47    & 2000 & -1.5 & 0.20  &  2.2   & 0.15  & 5.3  & 0.10  & 0.48  & 0.21  & 6.6 \\
   && R2 &  9500 &   47    & 2000 & -1.5 & 0.40  &  3.5   & 0.17  & 5.1  & 0.15  & 0.50  & 0.30  & 4.4 \\
   && R3 &  9500 &   47    & 2000 & -1.5 & 0.80  &  7.9   & 0.20  & 4.9  & 0.18  & 0.50  & 0.36  & 2.3 \\\hline
$H_{\innn}$ &
\ref{fig-stmodel-effect-all}C &
      H1 &  9500 &   47    & 2000 & -1.5 & 0.52  &  6.0   & 0.18  & 5.0  & 0.16  & 0.49  & 0.34  & 3.4 \\
   && H2 &  9500 &   47    & 2000 & -1.5 & 0.52  & 17.0   & 0.22  & 6.0  & 0.20  & 0.48  & 0.43  & 1.9 \\
   && H3 &  9500 &   47    & 2000 & -1.5 & 0.53  & 35.0   & 0.26  & 7.0  & 0.25  & 0.49  & 0.50  & 1.2 \\
   && H4 &  9500 &   47    & 2000 & -1.5 & 0.54  & 70.0   & 0.30  & 8.0  & 0.29  & 0.48  & 0.60  & 0.8 \\\hline
$\Sigma$ power &
\ref{fig-stmodel-effect-all}D &
      S1 &  9500 &   47    &    5 & -0.5 & 0.50  &  2.9   & 0.12  & 3.5  & 0.11  & 0.48  & 0.23  & 4.8 \\
   && S2 &  9500 &   47    &  100 & -1.0 & 0.52  &  4.2   & 0.16  & 4.3  & 0.14  & 0.48  & 0.29  & 3.9 \\
   && S3 &  9500 &   47    & 2000 & -1.5 & 0.52  &  6.0   & 0.18  & 5.0  & 0.16  & 0.49  & 0.34  & 3.4 \\
   && S4 &  9500 &   47    & $4\!\cdot\!10^4$ & -2.0 & 0.52  &  6.0   & 0.20  & 5.6  & 0.19  & 0.51  & 0.37  & 3.2 \\\hline
ZAMS &
\ref{fig-zams-seds} &
      Z1(B2) & 22000 & 1000    & 2000 & -1.5 & 5.88  & 68.0   & 0.27  & 4.5  & 0.31  & 0.51  & 0.60  & 2.6 \\
   && Z2(A0) &  8900 &  50     & 2000 & -1.5 & 0.53  &  6.1   & 0.18  & 5.0  & 0.16  & 0.48  & 0.34  & 3.4 \\
   && Z3(G2) &  5900 &   1     & 2000 & -1.5 & 0.08  &  0.73  & 0.12  & 5.4  & 0.10  & 0.48  & 0.21  & 3.3 \\
   && Z4(M2) &  3500 &   0.05  & 2000 & -1.5 & 0.015 &  0.22  & 0.09  & 5.7  & 0.07  & 0.48  & 0.15  & 2.2 \\\hline
T Tauri &
\ref{fig-tt-shadow} &
      T1 &  3800 &   0.76  & 2000 & -1.5 & 0.064 &  0.64  & 0.14  & 5.4  & 0.13  & 0.60  & 0.21  & 3.0 \\\hline
AB Aur &
\ref{fig:AB-Aur} &
      A1 &  9520 &  48.3   & 10$^4$ & -2.0 & 0.52  &  8.1   & 0.19  & 5.4  & 0.22  & 0.45  & 0.49  & 2.5 \\\hline
\multicolumn{16}{p{18.6cm}}{\revtwo{The first three columns are for model identification, the
next seven list model parameters and disk properties, and the last four
columns list computed observables evaluated from the star-subtracted
SED, where $F_{\mathrm{NIR}}$ is the
flux in interval 1.25--7 \um, $F_{\mathrm{IR}}$ the flux at $\lambda > 1.25$
\um, and $F_{10}$ and $F_{7.7}$ the fluxes at 10\um{} and 7.7\um,
respectively. $F_{*}$ is the integrated stellar flux.}}
\end{tabular}}}
\end{table*}

\section{Results}

\CH{In this section we show the structure of the computed models and
the SEDs produced by models with different parameters.}

\subsection{Example model}\label{subsec-standard-example}

To illustrate the general features of our model we set up an example
model consisting of a star \revone{with $T_{\eff}=9500\kel$},
$L_{*}=47\;L_{\odot}$ and $M_{*}=2.4\;M_{\odot}$ (we take the stellar
spectrum to be a blackbody for simplicity), and a disk with outer
radius $R_{\outtt}= 400\;\AU$ and inner radius at the dust
condensation point ($\trim=1500\kel$), which for this setup is at
$\rinn=0.47\AU$. \revone{The disk's surface density goes as
  $\Sigma(R)=2\times 10^3\;(R/\AU)^{\beta}$, with $\beta=-1.5$ and we
  assume a gas to dust ratio of 100}. The disk is seen at an
inclination angle $\iincl=45^o$. We take a simple model for our dust
opacity: a mixture of amorphous carbon and astronomical silicate in a
ratio $0.05$. The amorphous carbon opacity is computed according to
the simple recipe given by Ivezic et al (1997), while the silicate
opacity is that of Draine \& Lee (1984).  We ignore scattering.
\CH{The SED depends on dust properties in various ways, and
  variations of the opacity from object to object and between
  grains in the disk surface layer and midplane are known to occur
  (e.g.~Bouwman et al.  2000).  However, this aspect of the problem is
  outside the scope of this paper, and is not relevant to the
  conclusions we reach.}

\begin{figure}
\epsscale{0.5}
\plotone{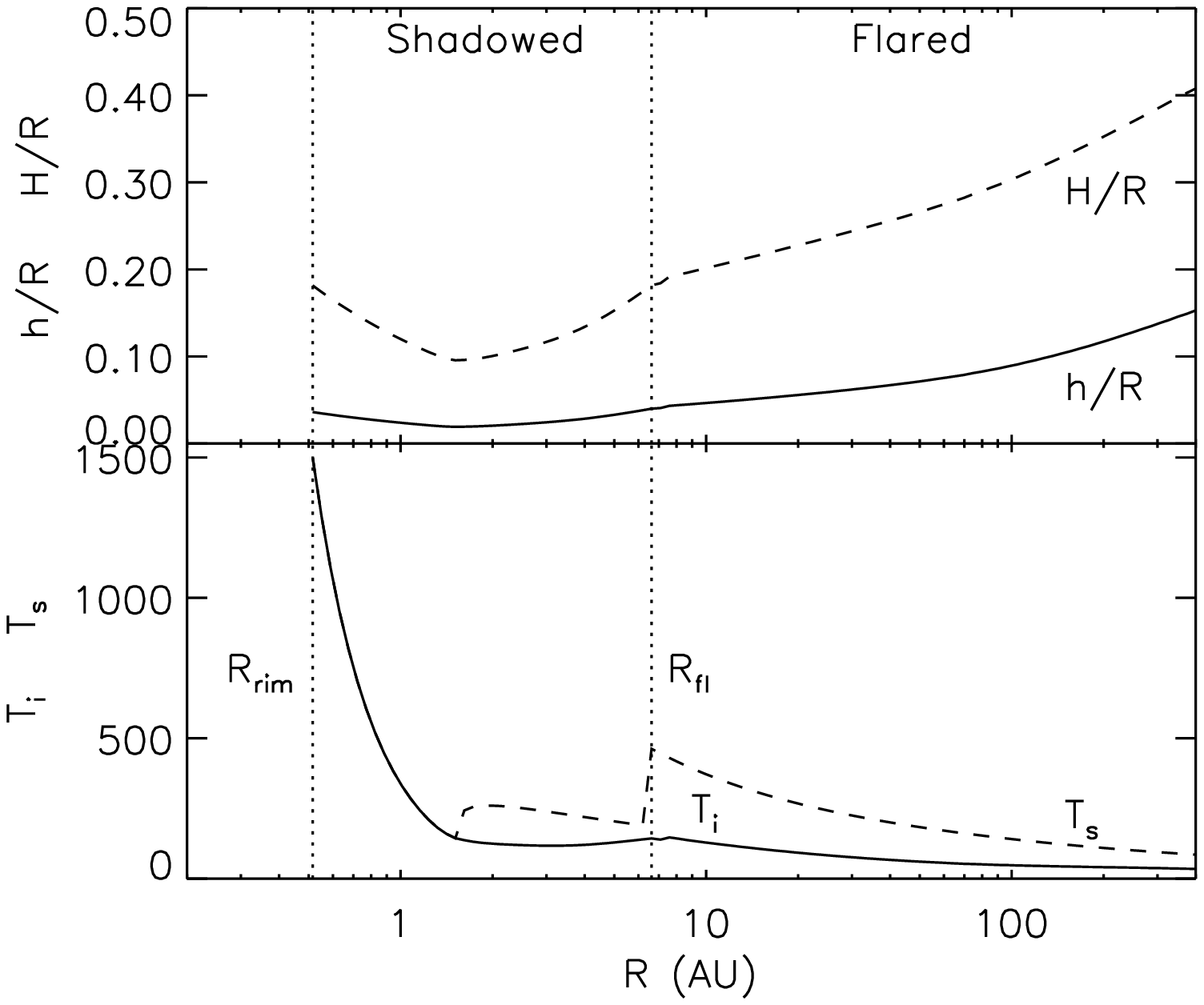}
\epsscale{1.0}
\caption{\revtwo{The structure of example model E3.  Top panel: disk
pressure scale height (solid line) and surface height (dashed line), both
divided by radius. Bottom panel: midplane temperature (solid line) and
temperature of the surface layer (dashed line).}}
\label{fig-computed-disk-structure}
\end{figure}

Fig.~\ref{fig-computed-disk-structure} shows the structure of the disk for
these stellar parameters.  The inner edge \revone{is at $R=0.52\AU$}, beyond
which the vertical height of the inner edge quickly declines (although when
plotted linearly in $R$ this decline is not so rapid as it seems in this
figure). \revone{For $R>6.6\AU$} the disk has the usual flaring shape
\revrm{predicted by CG97}, but with self-irradiation effects added. The
region \revone{between $0.52\AU<R<6.6\AU$} is the shadowed
region. \revone{Between $0.52\AU<R<1\AU$} the disk is dominated by radiative
diffusion from the inner edge outwards, as described in
Section~\ref{sec:shadowed-region}.  In the \revone{region $1\AU<R<1.5\AU$}
radiative diffusion from outside inwards dominates.  For \revone{radii
$1.5\AU<R<6.6\AU$} the disk is still not in sight of the star itself, but it
receives flux from the inner rim, which is sufficiently strong to keep up
the disk. Inwards radiative diffusion prevents a sudden jump to the fully
flared region \revone{starting at $R>6.6\AU$}, although the surface
temperatures cannot be prevented from jumping. The inward radiative
diffusion is estimated using an equation similar to Eq.~\eqref{eq:1}, but 
\revrm{taking into account the changing gravitational potential}
\revone{with opposite sign}. The
details of the structure \revone{between $1\AU<R<6.6\AU$} are not very
important, since this region hardly contributes to the SED. For this reason
we are content with the rather crude description presented here.

\begin{figure}
\plotone{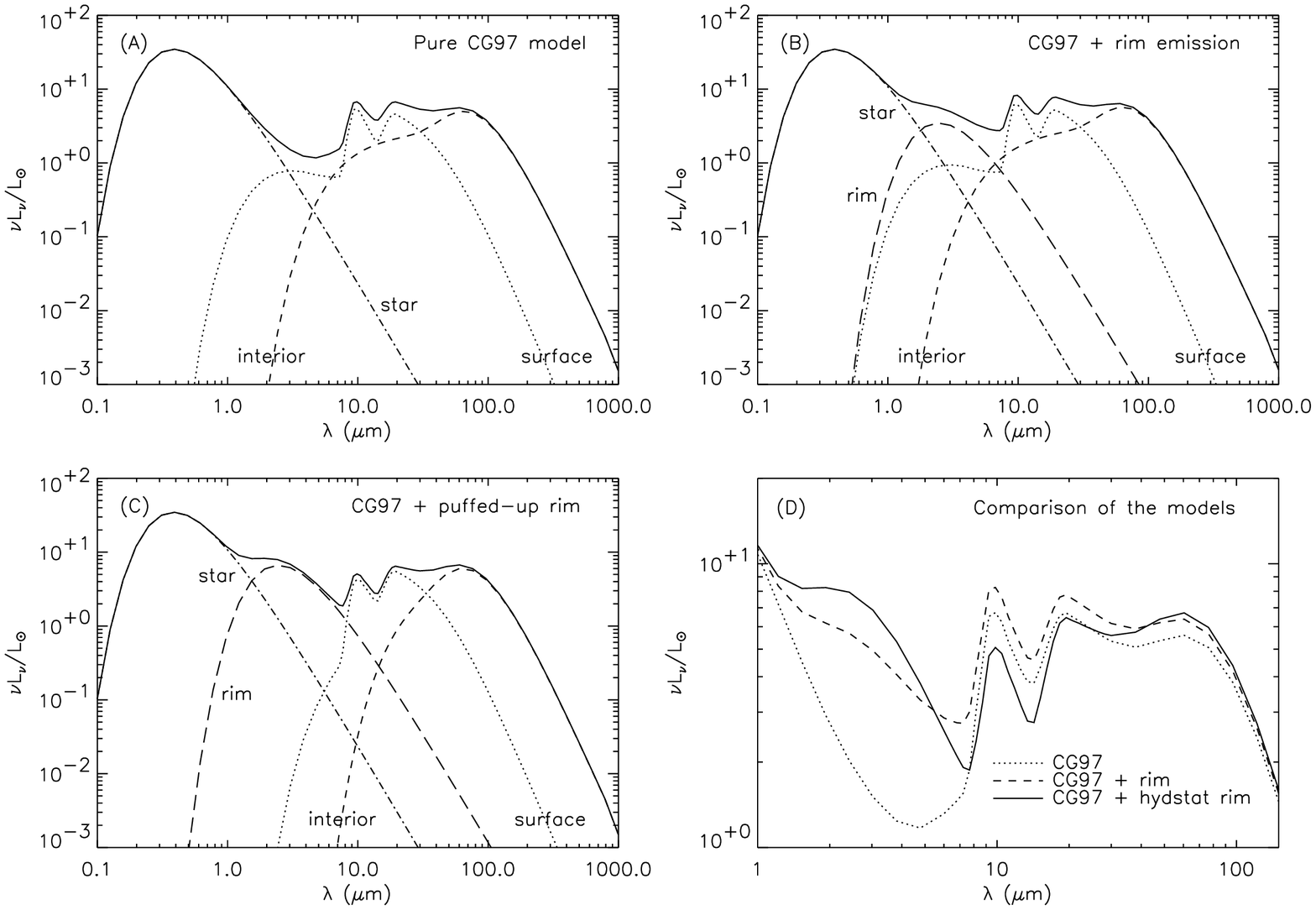}
\caption{\label{fig-stmodel-sed-all} SED of the example model (see section
\ref{subsec-standard-example}), computed according to the unmodified
\CH{CG97 model (panel A, model code E1), the CG97 model with inclusion of
the emission from the inner rim (panel B, model code E2), and with a
self-consistent puffed-up inner rim (panel C, model code E3)}. In panel D
the three SEDs are compared. In the unmodified CG97 model, the SED (solid
line) is the sum of the emission of the star (dash-dotted line), the disk
interior (dashed line) and the disk surface (dotted line).  When the
emission from the inner edge is added, without modifying the vertical height
of the inner rim, then a blackbody component at NIR wavelengths is added to
the SED (long-dashed line in Panel B). When the height of the inner rim is
consistently computed, the SED is also the sum of four components (same
symbols as in Panel B), but the NIR emission increases, and a shadow is cast
over the disk, thus suppressing some of the disk emission.}
\end{figure}

Fig.~\ref{fig-stmodel-sed-all} shows the resulting SED for the example
model. We show for reference in panel (A) the SED predicted by a CG97 model,
where the effects of the inner radius (emission and shadow) is ignored.
Panel (B) shows the effect of including the emission from the inner rim,
assumed to have the height of \revone{the flaring} disk at that radius.  In this case, the
SED in the mid to far infrared remains unmodified, but a new component
appears in the near-infrared.  Finally, panel (C) shows the results of the
self-consistent calculation described in the previous sections.  One can see
that the NIR emission from the puffed-up rim increases, but the inner rim
starts to cast a shadow over the disk, thereby suppressing some of the mid-
and far infrared emission. Panel (D) shows a blow-up of the 1--100 \um\
region, where the different effects of the inner rim can be clearly seen.

\revone{In the following subsections we will show how the model-predicted SEDs
change as function of some of the model parameters. We summarize some
of the characteristics of the models and of the SED properties in
Table~1. }
%

\subsection{Effect of inclination angle}
\begin{figure}
\epsscale{0.5}
\plotone{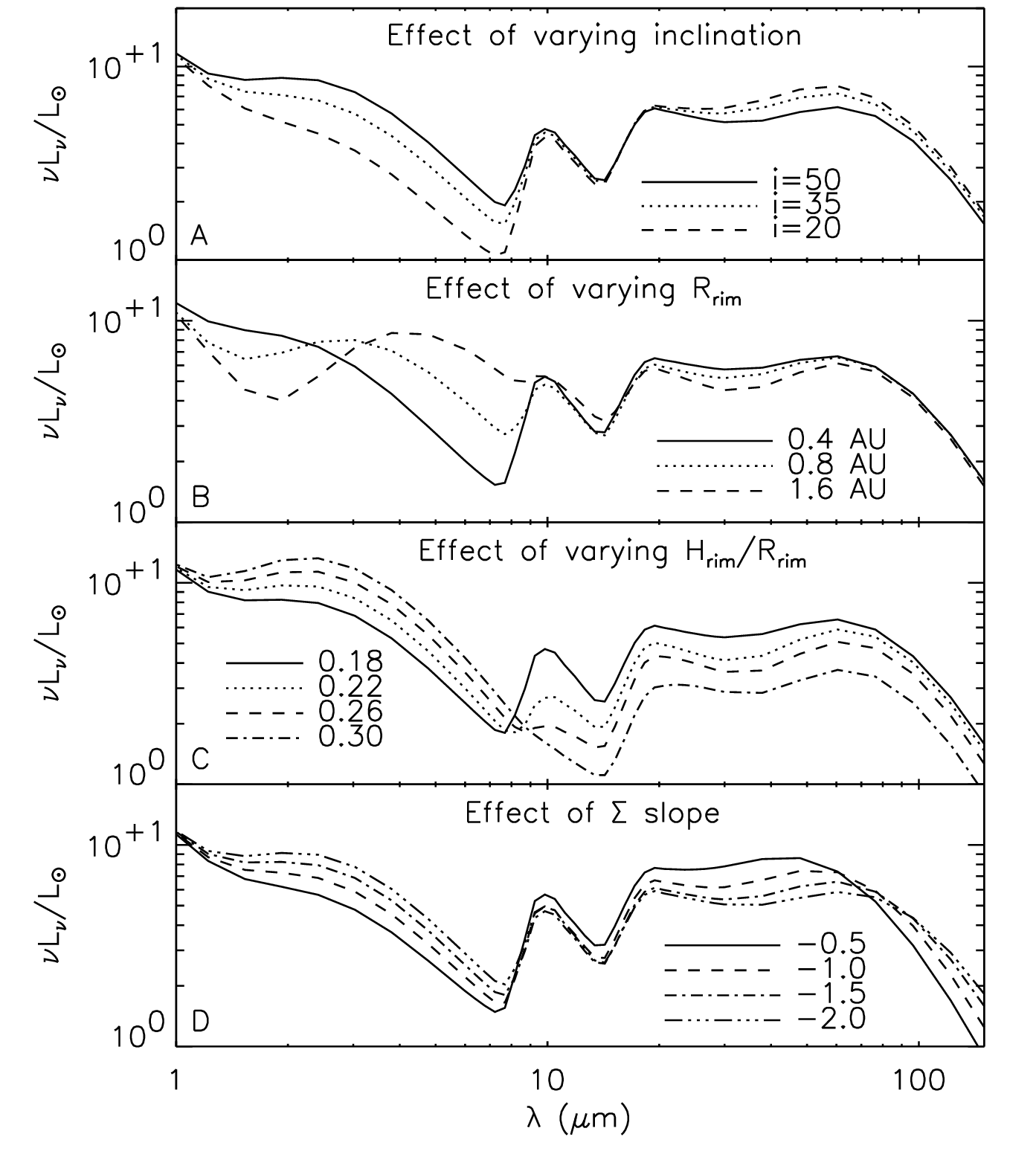}
\epsscale{1.0}
\caption{The observed SED of our example model, for varying input
parameters. For all figures the inclination angle is $\iincl=45$ degrees,
except for the one that shows the effect of varying inclination.}
\label{fig-stmodel-effect-all}
\label{fig-stmodel-sed-incl}
\label{fig-stmodel-rin}
\label{fig-ten-mic-feature}
\label{fig-stmodel-sigmaslope}
\end{figure}

In Fig.~\ref{fig-stmodel-sed-incl} we show how the SED of our example model
changes with the inclination with respect to the line of sight (Panel A). We
consider here only values of $\iincl$ such that the line of sight does not
intercept the outer disk.

One sees that for nearly face-on inclinations the NIR excess is less
prominent and the far infrared is strong, while for nearly edge-on
inclinations the opposite is true. By varying $\iincl$ from $50^o$ to
$20^o$, the ratio \rnir\ decreases from 0.18 to 0.08, while the infrared
excess at wavelengths longer than 7 \um\ increases from 0.32 to 0.35.  This
is because the disk radiates predominantly towards the poles, while the
inner rim radiates predominantly towards the equator. The emission features
are optically thin features and are therefore less affected by the
inclination.  Only when the inclination would exceed the opening angle of
the flared disk, these features would go from emission into absorption
because of the obscuration effect of the disk itself. However, in order to
model this properly we would need a multidimensional ray-tracing
computation, which is beyond the scope of this paper.

\subsection{Effect of varying inner radius}
The inner rim in our models is located at the dust evaporation radius, and
is therefore fixed once the stellar parameters are defined.  Nevertheless,
it is instructive to see the effects of varying inner radius on the SED,
because it helps in clarifying the properties of our models, and we plot in
Fig.~\ref{fig-stmodel-rin} the SED of our standard model for varying $\rinn$
(Panel B). \CH{The effect is strong in the near-infrared, where both the
excess luminosity and its shape change drastically. For larger $\rinn$, the
near-infrared ``bump'' is enhanced, with a peak that moves toward longer
wavelengths, but the flux around 1\um\ decreases as the temperature of the
inner rim decreases.}

\CH{There are also, barely visible in the figure, some changes at
intermediate wavelengths, due to the reduction of shadowing which
increases the contribution of the disk surface layer also around 50\um.}

\subsection{Effect of shadow on the 10 micron feature}
If the height of the puffed-up inner rim is computed according to the
equation for vertical pressure balance (Eq.~\eqref{eq-vert-press-bal-cg}
with $\Ti$ replace by $\trim$), then the shadow that the rim projects over
the disk will usually extend up to a radius of roughly $10$ to $20$ times
the inner radius. This ratio turns out to be fairly robust in our model,
which can be attributed to the fact that the ratio of $\trim/\tdisk$ is
fairly constant for all model parameters. At $10$ to $20$ times $\rinn$ the
surface layer temperature is about $4.5$ to $6.3$ times lower than $\trim$
(which is $1500\kel$ by definition in our model).  This means that the
shadow starts to affect the regions of the disk's surface layer that emit
the 10 micron feature, which lies at $5$ times longer wavelengths than the
NIR bump. This effect is already visible when comparing our standard model
to the result of a CG97 model (Fig.~\ref{fig-stmodel-sed-all}C and D).

If, for some reason, we have underestimated the height of the inner rim
(which is the most difficult quantity to determine in our models), this
might have repercussions on the predicted 10 micron flux.  To investigate
this effect, we have artificially enhanced the height of the puffed-up inner
rim (i.e.~we have relaxed the assumption that the disk is here in vertical
pressure balance), and computed the SED. The results are shown in
Fig.~\ref{fig-ten-mic-feature}C, where we have increased $\chi_{\innn}$ to
values as large as 8.  For increasing inner rim height, the 2 micron bump is
of course enhanced, but the major effect on the SED is the change at longer
wavelengths, due to the fact that the shadow extends to larger radii,
eventually to most of the disk. The 10 micron feature is the first part of
the SED to be affected by the increase of $\chi_{\innn}$, and has entirely
disappeared for $\chi_{\innn}$=8.  The 20 micron silicate feature is less
affected, because it is emitted at lower temperature, and hence at larger
radius. The total IR flux is conserved: the increase in NIR flux goes at the
cost of the mid-IR and far-IR flux (see Table 1, models H1..H4).

The exact value of $\chi_{\innn}$ depends on a number of properties that are
difficult to constrain, such as the shape of the disk behind the rim and the
dust properties.  However, its exact value is not crucial for the discussion
of the basic idea behind our model, which revolves around two issues: the
presence of a puffed-up inner rim at roughly the dust evaporation radius,
naturally producing the NIR bump, and a proper consideration of energy
conservation, which implies that we must take into account the shadow the
rim projects onto the disk.  The exact shape of the rim is, in a way, a
secondary aspect.

\subsection{Effect of $\Sigma$-slope}
\revone{The surface density $\Sigma(R)$ is an unknown quantity, and serves
  as an input to the model. The results of the models turn out not to be
  very dependent on the surface density. The main effect it can have is
  changing the balance between near-IR and far-IR radiation, and the slope
  of the SED at far-IR/submm wavelengths.  To demonstrate this, we show
  models with $\Sigma=\Sigma_0(R/R_0)^\beta$ for different values of
  $\beta$. For each of these models we tune the $\Sigma_0$ such that the
  total covering fraction of the disk $\Omega=\Hs(\rout)/\rout$ is equal to
  $\Omega=0.4$, see table 1 for the resulting values of $\Sigma_{0}$. The
  rest of the parameters are the same as for the example model. The results
  are shown in Fig.~\ref{fig-ten-mic-feature}D.}

\subsection{Behavior of the model for varying stellar type}
\begin{figure}
\epsscale{0.5}
\plotone{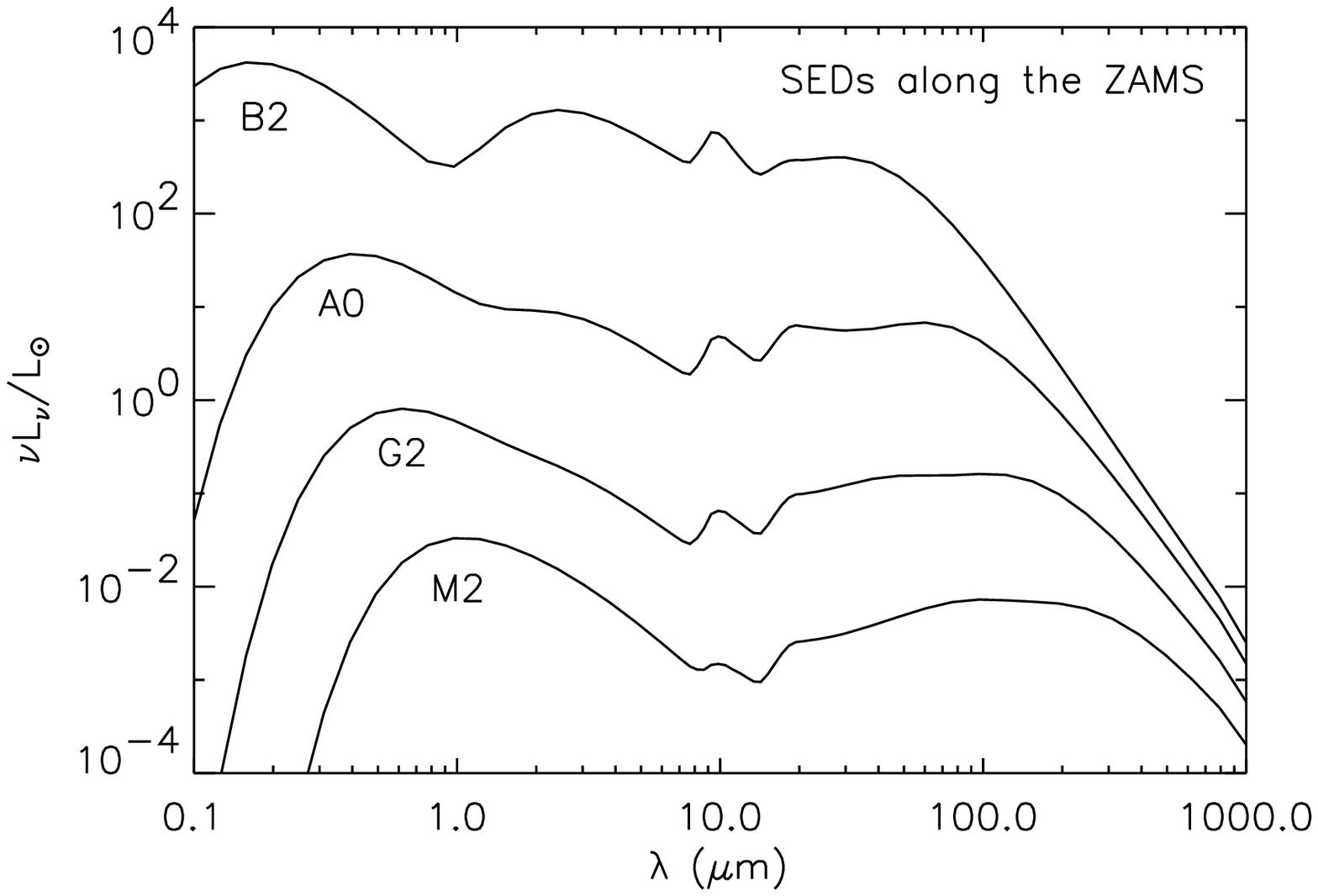}
\epsscale{1.0}
\caption{Model-predicted SED for  ZAMS stars of different
spectral type. The stellar parameters are given in Table 1.}
\label{fig-zams-seds}
\end{figure}

\begin{figure}
\epsscale{0.5}
\plotone{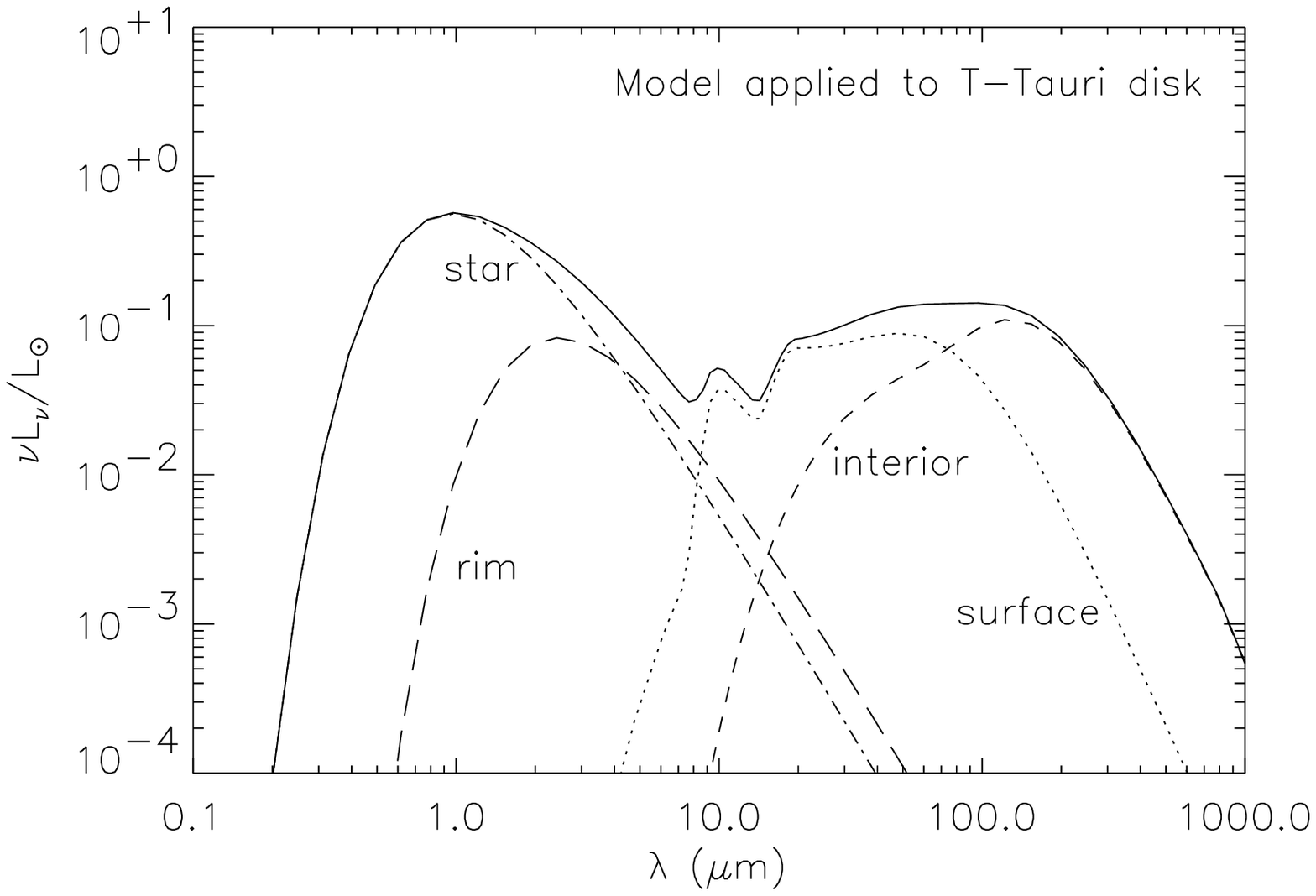}
\epsscale{1.0}
\caption{SED of a T Tauri star of $T_{\eff}=3800$ K, 
$L/L_{\odot}=0.76$ and $M/M_{\odot}=0.55$. \revtwo{The various line styles
represent the different components contributing to the SED in the same
way as in Fig.(\ref{fig-stmodel-sed-all}).}}
\label{fig-tt-shadow}
\end{figure}

The behaviour of the near-infrared excess as function of the stellar type of
the central star is shown in Fig.~\ref{fig-zams-seds}, which plots the SEDs
of four stars along the zero-age-main-sequence ranging from spectral type B2
to M2 (see Table 1).  The NIR bump is clearly visible in stars of spectral
type A and earlier, but tends to disappear in later stars.  This is due to
two effects. First of all, for cooler central stars the stellar flux peaks
at longer wavelength, and therefore swamps the near-infrared emission of the
disk. In addition to this, the ratio of the covering factor of the
puffed-up inner rim (in vertical hydrostatic equilibrium) to the covering
fraction of the flaring part of the disk decreases for later spectral
types. These factors determine the relative importance of the various
components of the SED, as can be seen from Table 1.  As we consider stars of
later spectral types, the near-infrared excess luminosity decreases by a
factor of two.  The various component that contribute to the observed SED of
a typical TTS are shown in Fig.~\ref{fig-tt-shadow}.

Our disk models provide a natural explanation for the fact that the
near-infrared ``bump'' is observed in Herbig AeBe stars, while it is not
noticeable in stars of later spectral type, and in particular in TTS. We
will come back to this point in the following section.


\section{Discussion}

\subsection{Disk physics}

\CH{It is currently not clear whether a sharp inner rim can be stable
  in the way described here.  The thermal pressure which is
  responsible for the rim height will act not only in the vertical but
  also radially toward the star.  We can think of two possibilities to
  stabilise the rim.  The space between the star and the disk could be
  filled with optically thin gas, and the gas pressure could stabilize
  the rim.  A second possibility would be centrifugal force.  If
  the inner rim rotates at slightly super-keplerian speeds, it could
  counter the gas pressure.  This super-keplerian rotation might be
  automatically created if material flowing inwards because of an
  initial pressure gradient preserves it angular momentum.  A
  dynamical study of the problems would be very interesting, but is
  beyond the scope of the current study.}

\revone{Radiation pressure on dust grains may also affect the
  structure of the inner rim.  In the upper regions, the gas drag may
  not be strong enough to keep small dust grains from being blown away
  by radiation pressure, an effect which would locally decrease the
  dust-to-gas ratio and therefore opacity and the rim height.  On the
  other hand, radiation pressure could also lead to gas-dust
  separation at the inner rim if some low level accretion is still
  going on in the disk.  The star would then accrete only the gas while
  the dust would stay in the rim and accumulate with time.  This might
  cause an \emph{increase} of the rim height, compensating for the
  removal of grains by radiation pressure.  Radiation pressure on dust
  grains can also have effects in the flaring part of the disk, by
  pressing the grains deeper into the disk and reducing the surface
  height there.  All these processes are poorly studied and deserve
  more attention in future work.}

\CH{Another point of concern is the question whether the flaring part of
the disk is dynamically stable. If one takes the disk equations of
CG97, and those described in the present paper, at face value, one can
show that the disk is unstable to self-shadowing effects caused by a
perturbation in the vertical scale height (Dullemond 2000, Chiang
2000). In fact, the shadowed region outside the inner rim is a an
example of this shadowing effect.  This instability acts on a
Keplerian time scale and could cause the entire flaring disk to
collapse into the shadow of the inner rim.  A recent study however
shows that multi-dimensional radiative transfer effects can stabilize
the disk for certain disk parameters (Dullemond submitted).  And the
inclusion of self-consistent vertical structure of the disk models may
enhance this stabilizing effect.}

\subsection {Comparison with the observations}

Disk models where the irradiation from the star dominates the heating
provide a good fit to most properties of pre-main-sequence stars over a
large range of masses \revone{(see, for example, D'Alessio et al.~1998,
1999, CG97, Chiang et al.~2001).} However, there are few points that have
remained unclear, and we think that the better treatment of the inner disk
with respect to previous models provides an answer to one of them, namely
the long-standing puzzle of the near-IR bump in the SEDs of HAeBe stars.

\begin{figure}
\epsscale{0.5}
\plotone{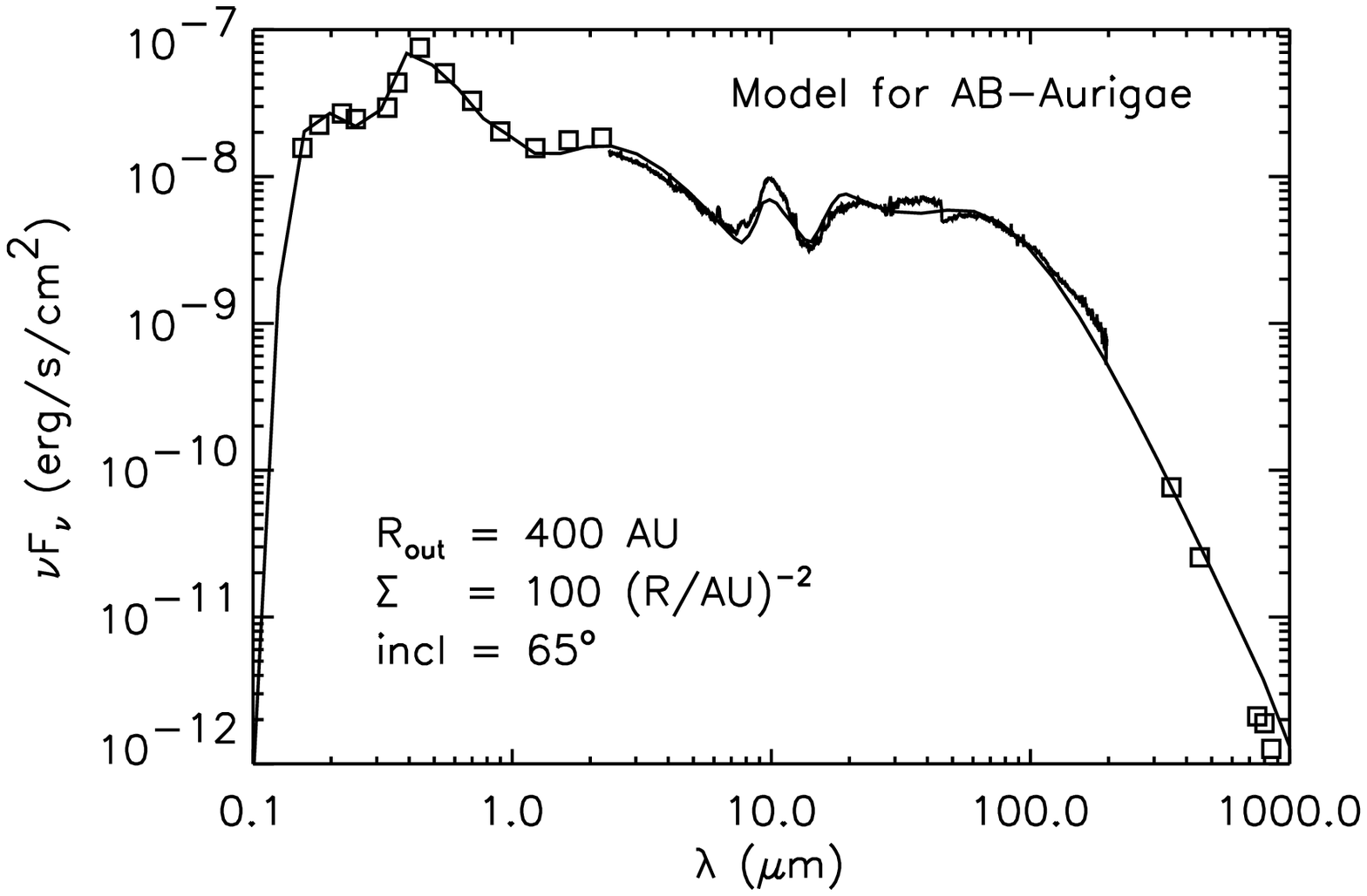}
\epsscale{1.0}
\caption{\CH{\revone{Best fit to AB Aur. Parameters are
$M_{*}=2.4M_{\odot}$, $L_{*} = 47L_{\odot}$, $T_{*}=9520$K, $\Sigma = 10^4
*(R/\mathrm{AU})^{-2}$, R$_{\mathrm{out}}=400$AU, $\iincl=65^{\deg}$,
$D=144$pc. \revtwo{The plot shows the model fit (solid line), together with
photometery points (squares) and the ISO SWS and LWS spectra between 2.3 and
200 $\mu$m (data from Meeus et al.~2001).}}}}
\label{fig:AB-Aur}
\end{figure}

In Fig.~\ref{fig:AB-Aur} we show a fit of our model to the SED of
AB~Aurigae, one of the best studied HAe stars. The basic features of the SED
are well reproduced by the model: the 3\um\ bump, the 10 micron feature, the
far IR plateau and the sharp decline towards mm wavelengths. Our model has
only a few free parameters ($i$, $\Sigma(R)$, $\rout$ and the dust opacity
table), so the fact that the data can be fit in a reasonable way gives
support to our model. The inclination angle of $\iincl=65^{\deg}$ needed for
this fit is larger than what most observations suggest (e.g.~Natta et
al. 2001 and references therein).  However, the assumption of a perfectly
vertical rim which is responsible for this value may be too simple.  A more
rounded surface of the inner rim will allow more of the rim radiation to be
emitted into polar directions.

Nevertheless, the model-predicted values of \rnir\ for A stars (hovering
around 0.17) agree well with those observed in four HAe stars by Natta et
al.~(2001), which range between 0.12 and 0.25, respectively, and by Meeus et
al.~(2001) in a sample of 14 HAe stars, where \rnir\ varies between 0.06 and
0.3 for the group I sources.

Observations show that the ``bump'' is a more or less constant feature from
star to star (Hillenbrand et al. 1992), in contrast to the long-wavelength
emission, which is affected by a variety of disk and dust parameters (Meeus
et al. 2001). This robustness is naturally reproduced in our model, since to
first order the properties of the NIR bump are determined only by the dust
sublimation temperature and the covering fraction of the rim, the latter
varying only weakly as a function of stellar type.

In spite of this, the ``bump'' disappears in stars of spectral type F or
lower (see Fig.~\ref{fig-zams-seds}). As discussed in \S 4.6 and shown in
Table 1, this can be explained as being due to two concurring effects. The
first is the lower covering angle of the rim in stars of later spectral
type, where dust sublimation occurs closer to the star. The second is the
fact that the stellar radiation peaks at longer wavelengths, swamping the
rim emission. Thus, even if the shape of the ``bump'' is the same, it
becomes very difficult to see it and to separate the rim contribution from
the stellar emission. It is, however, possible that one is seeing the rim
emission in some TTS with near-infrared colors that can only be understood
if the disk has an inner radius of few stellar radii, coinciding with the
dust sublimation radius for systems of typical TTS luminosity (Beckwith et
al. 1990; Kenyon et al. 1996).

Another interesting aspect of these models is the effect of the shadow that
the rim projects onto the disk. The 10 \um\ silicate emission in our example
model has \fsil=3.5, in reasonable agreement with what observed in HAe stars
by Natta et al. (2001). These numbers, however, are very dependent on the
properties of the grains in the disk surface layer. A weak or absent 10 \um\
feature, for example, can be attributed to a dominance of large silicate
grains in the disk surface layer (Meeus et al. 2001), to a very high ratio
of the visual opacity to the opacity in the feature (Natta et al. 2001), or
to an almost perfect cancellation of absorption and emission in disks seen
edge-on (Chiang \& Goldreich 1999).  We propose here still another
possibility, namely that the inner rim in some object is higher than normal,
because, for example, the rim is not in hydrostatic equilibrium, is thicker
in the radial direction, or has a higher $\chi_{\innn}$, the ratio of the
photospheric to the pressure scale height. As shown in
Fig.~\ref{fig-stmodel-effect-all}C and Table 1, a factor 1.7 increase of
$\chi_{\innn}$ would be sufficient to suppress the silicate feature
entirely.

One point we want to stress here is that a rim structure such as we have
described can only exist if the gaseous disk is optically thin to the
stellar radiation {\it and} the dusty disk is optically thick. If an
optically thick gaseous disk extends toward the star closer than the dust
evaporation radius, then there is no puffed-up rim and the standard models,
which describe the emission of an optically thick disk surface, apply. The
formation of a puffed-up rim seems possible only if the accretion rate
through the disk is very low, as briefly discussed by Natta et
al. (2001). There are indications that this is indeed the case in the
majority of HAe stars.  Tambovtseva et al. (2001) find accretion rates below
$10^{-8}$ M$_\odot$ yr$^{-1}$ from their study of Balmer emission lines in
UX Ori (see also Ghandour et al. 1994).  There are no similar studies for
the hotter HBe stars, and it is clear that in many TTS accretion rates are
high enough to ensure that the inner gaseous disk will be optically thick
(Gullbring et al. 1998).  This is an additional explanation why the
near-infrared ``bump'' is seen only in a minor fraction of TTS, if at all.

The idea that the emission in the near-infrared is dominated by the emission
of the inner rim finds support in the recent interferometric observations of
Millan-Gabet et al. (2000). They find that in J and K the emission of AB~Aur
is best fitted by the emission of a thin shell of radius $\sim 0.3$ AU,
almost spherical on the plane of the sky. For the luminosity of AB~Aur, the
inner rim is located at 0.5 AU for dust evaporation temperature around 1500
K. For the low inclination derived from millimeter and optical data ($\iincl
\simless 45^o$; see Natta et al. 2001 and references therein), the rim
projected on the plane of the sky is almost spherical.  We expect that other
HAe stars will be resolved by current and future interferometers, so that
the main feature of our models, namely the fact that the near-infrared
emission comes from a narrow rim located at the dust evaporation radius,
will be directly checked.

%
%
\begin{acknowledgements}
%
%
%
%
\revone{We would like to thank the anonymous referee for a careful review of
  the paper. CD acknowledges the finantial support fomr NWO Pionier grant
  6000-78-333.  CPD and CD acknowledge support from the European Commission
  under TMR grant ERBFMRX-CT98-0195 (`Accretion onto black holes, compact
  objects and prototars'). We thank Rens Waters, Jeroen Bouwman, Caroline
  Terquem, Claude Bertout, Alex de Koter, Vincent Icke and Rohied Mokiem for
  interesting discussions, and Gwendolyn Meeus for access to her
  observational data of HAeBe stars.}
%
%
\end{acknowledgements}
%
%
%
%

\pagebreak[4]

\pagebreak[4]

\appendix

\section{Fine-tuning functions}
\label{app:sec-cg-finetune}

\subsection{The $\psi$ parameters}
\label{app:subsec-cg-finetune-psi}
When the interior of the disk is not optically thick anymore to its own
radiation and/or the radiation from the surface layer, then $\psi$ will
deviate from unity. For an optically thick interior, the total emitted flux
(one a single side) is simply $F_{\emit}=\sigma \Ti^4$. For non optically
thick cases this becomes
\begin{equation}
\begin{split}
F_{\emit} &= \int_0^\infty\pi B_\nu(\Ti) \;[1-\exp(-\Sigma\kappa_\nu)]d\nu\\
&=\psi_{\interior}\;\sigma\Ti^4
\fullstop
\end{split}
\end{equation}
with
\begin{equation}
\psi_{\interior} = \frac{\int_0^\infty B_\nu(\Ti) 
\;[1-\exp(-\Sigma\kappa_\nu)]d\nu}{
\int_0^\infty B_\nu(\Ti)d\nu}
\end{equation}
The limits for high and low optical depth are
\begin{equation}\label{eq-limit-psi-int}
\psi_{\interior} = \left\{\quad 
\begin{matrix}
1 \quad & \hbox{for} \quad \Sigma\kappa_P(\Ti) \gg 1 \\
\Sigma\kappa_P(\Ti) \quad & \hbox{for} \quad \Sigma\kappa_P(\Ti) \ll 1 
\end{matrix}
\right.
\end{equation}

The absorbed flux from the surface layer equals the flux emitted downwards
by the surface layer multiplied by the absorption fraction
$[1-\exp(-\Sigma\kappa_\nu)]$, and integrated over frequency:
\begin{equation}
\begin{split}
F_{\abs} &= \int_0^\infty 2\pi\Delta\Sigma\,\kappa_\nu 
B_\nu(\Ts) \;[1-\exp(-\Sigma\kappa_\nu)]d\nu\\
&= \psi_{\surf}\;F_{\surf}
\end{split}
\end{equation}
where $F_{\surf}$ is the total flux from the surface layer downwards, and
\begin{equation}
\psi_{\surf}=\frac{\int_0^\infty B_\nu(\Ts) \kappa_\nu 
\;[1-\exp(-\Sigma\kappa_\nu)]d\nu}{
\int_0^\infty B_\nu(\Ts)\kappa_\nu d\nu}
\end{equation}
The limits for high and low optical depth are
\begin{equation}\label{eq-limit-psi-surf}
\psi_{\surf} = \left\{\quad 
\begin{matrix}
1 \quad & \hbox{for} \quad \Sigma\kappa_P(\Ts) \gg 1 \\
\Sigma\kappa_Q^2(\Ts)/\kappa_P(\Ts) \quad & \hbox{for} \quad \Sigma\kappa_P(\Ts) \ll 1 
\end{matrix}
\right.
\end{equation}
where $\kappa_Q$ is the Planck square mean opacity:
\begin{equation}
\kappa_Q^2(T)\equiv \frac{\int_0^\infty B_\nu(T)\kappa_\nu^2 d\nu}
{\int_0^\infty B_\nu(T) d\nu}
\end{equation}
To a certain degree of accuracy the approximation
$\kappa_Q^2(\Ts)/\kappa_P(\Ts) \simeq \kappa_P(\Ts)$ can be used.

When the above limiting expressions for $\psi_{\surf}$
(Eq.\ref{eq-limit-psi-surf}) and $\psi_{\interior}$
(Eq.\ref{eq-limit-psi-int}) are inserted into Eq.(\ref{eq-cgeq-tempi}),
then one reproduces Eqs.(12a,12b,12c) of CG97.

Since these $\psi_{\interior}$ and $\psi_{\surf}$ depend on $\Ti$ itself
(the determination of which requires the values of $\psi_{\interior}$ and
$\psi_{\surf}$), the solution of Eq.(\ref{eq-cgeq-tempi}) requires an
iterative procedure.

\subsection{The $\chiCG$ parameter}
\label{app:subsec-cg-finetune-chi}
For a given grazing angle $\alpha$, the ratio $\chiCG\equiv \Hs/\Hp$ is
independent of $\Hp$. However, $\alpha$ itself depends on $\Hs$ through
Eq.(\ref{eq-cgalpha}). We define $\Hs$ to be the height at which
\begin{equation}
\frac{\kappa_P(T_{*})}{\alpha} \int_{\Hs}^\infty \rho(z) dz = 1
\end{equation}
Since the vertical density distribution is assumed to be a Gaussian, this
amounts to solving the following equation
\begin{equation}\label{eq-solve-chi}
1-\erf(\chiCG/\sqrt{2}) = \frac{2\alpha(\chiCG)}{\Sigma\kappa_P(T_{*})}
\comma
\end{equation}
where, for a given value of $\Hp$, the $\alpha$ depends on $\chiCG$ through
Eq.(\ref{eq-cgalpha}). Eq.(\ref{eq-solve-chi}) can be quickly solved for
$\chiCG$ using any kind of root-finding algorithm.

\subsection{The $\chiin$ parameter}
\label{app:sec:surface-height-at-inner-rim}
In this appendix we wish to determine the dimensionless parameter
$\chiin=\Hsin/\Hpin$, which determines the vertical height of the
inner rim. We define the surface height to be the height to which the
optical depth of the rim on a radially outward directed ray is greater
than 1.
\begin{equation}
\tau_{\radial}(z_0) = \int_{\rinn}^\infty \rho(R,z) \kappa_P(T_{*}) dR = 1
\comma
\end{equation}
where $z = z_0 R$ and $z_0$ is the value of $z$ at $R=\rinn$. \revone{We
neglected a geometric factor $\sqrt{1+z_0^2/\rinn^2}$ (arising from
cylindrical coordinates) because it is very close to unity.}

In order to determine $\tau_{\radial}(z_0)$ we need to know the
two-dimensional structure of the puffed-up inner rim, which we discussed in
section~\ref{sec:shadowed-region}. We have estimated the radial behavior of
$H(R)/R$ to be linear with a slope of $-1/8$. \revone{A radial ray through
the upper layers of the inner edge will therefore have a roughly 8 times
higher optical depth than a vertical ray between $z=z_0$ and $z=\infty$. In
order to estimate $z_0$ we therefore estimate $\tau_{\radial}(z_0)$ to be
simply 8 times $\tau_{\vert}(z_0)$:
\begin{equation}
\begin{split}
\tau_{\radial}(z_0) &= 8 \int_{z}^\infty \rho(\rinn,z)\kappa_P(T_{*})
dz' \\
&= \frac{8\Sigma\kappa_P(T_{*})}{\sqrt{\pi}}\int^{\infty}_{z/\Hpin} e^{-x^2} dx
\comma
\end{split}
\end{equation}
which can be easily computed following the procedure in appendix
\ref{app:subsec-cg-finetune-chi}.}

\section{Emission from a cylinder}\label{app:projsurf}

\subsection{Far field limit}
Consider the inner rim to be a cylinder with radius $R_{\innn}$ and
vertical height $\Hsin$, emitting toward the inside as a blackbody
of temperature  $T_{\innn}$. The unocculted surface as seen by an
observer at inclination $\iincl$ (measured from the pole) can be computed
as follows. Define the quantity $\dlt$:
\begin{equation}
\dlt \equiv \frac{\Hsin}{R_{\innn}}\tan \iincl
\fullstop
\end{equation}
$\dlt$ is defined so that
for $\delta=1$ the inclination is just small enough that we are able to
see the central star (which is assumed to be a point-source). For $\delta<1$
we have a more face-on view, while $\delta>1$ means a more edge-on view
of the system. The observed flux at a distance $d$ from this cylinder
in the far-field limit is for $\dlt<1$
\begin{equation}\label{eq-flux-from-rim-incl-dsmall}
F_\nu = 2 B_\nu(T_{\innn})\left(\frac{R_{\innn}}{d}\right)^2 \cos \iincl \left[
\dlt\sqrt{1-\dlt^2}+\arcsin\dlt\right]
\comma
\end{equation}
and for $\dlt > 1$
\begin{equation}\label{eq-flux-from-rim-incl-dlarge}
F_\nu = \pi B_\nu(T_{\innn})\left(\frac{R_{\innn}}{d}\right)^2 \cos \iincl 
\fullstop
\end{equation}

\subsection{Near field limit}

In the far-field limit, the rim becomes invisible at $\iincl=\pi/2$ which
corresponds to $z/R=0$.
In order to compute the contribution of the inner rim to the
irradiation of the flared disk, we need to correct the equation for
the fact that the rim becomes invisible already at $z=\Hsin$. We do
this by introducing a modified inclination $\theta$ defined by
\begin{equation}
\tan \theta = \frac{R-R_{\innn}}{z-\Hsin}
\end{equation}
where we took the reference point to be the nearest corner of the
cylinder instead of the star.  We can then write $\delta$ as
\begin{equation}
\delta = \frac{R/R_{\innn}-1}{z/\Hsin-1}
\end{equation}
The flux is then further given by
Eqs.(\ref{eq-flux-from-rim-incl-dsmall},\ref{eq-flux-from-rim-incl-dlarge}).
Comparing the resulting flux with exact numerical evaluation of the
projected surface has shown that this approximation gives irradiation
fluxes accurate to within 20\% for inclinations $>\pi/4$ and
$R>2R_{\innn}$.

\end{document}